\documentclass[sn-basic]{sn-jnl}

\usepackage{subfigure}
\usepackage{array}

\usepackage{graphicx}%
\usepackage{multirow}%
\usepackage{amsmath,amssymb,amsfonts}%
\usepackage{amsthm}%
\usepackage{mathrsfs}%
\usepackage[title]{appendix}%
\usepackage{xcolor}%
\usepackage{textcomp}%
\usepackage{manyfoot}%
\usepackage{booktabs}%
\usepackage{algorithm}%
\usepackage{algorithmicx}%
\usepackage{algpseudocode}%
\usepackage{listings}%
\usepackage{float}
\usepackage{array}

\begin{document}

\title[]{How cancer emerges: Data-driven universal insights into tumorigenesis via hallmark networks}


\author[1]{\fnm{Jiahe} \sur{Wang}}
\equalcont{These authors contributed equally to this work.}

\author[1]{\fnm{Yan} \sur{Wu}}
\equalcont{These authors contributed equally to this work.}

\author[1]{\fnm{Yuke} \sur{Hou}}

\author[1]{\fnm{Yang} \sur{Li}}

\author[1]{\fnm{Dachuan} \sur{Xu}}

\author*[1]{\fnm{Changjing} \sur{Zhuge}}\email{zhuge@bjut.edu.cn}

\author*[2]{\fnm{Yue} \sur{Han}}\email{hy19870705@sina.com}

\affil[1]{\orgdiv{Institute of Operations Research and Information Engineering}, \orgname{Beijing University of Technology}, \orgaddress{ \city{Beijing}, \postcode{100124}, \country{China}}}

\affil[2]{\orgdiv{Department of Gynecology}, \orgname{Affiliated Hospital of Nanjing University of Chinese Medicine}, \orgaddress{ \city{Nanjing}, \postcode{210029}, \state{Jiangsu}, \country{China}}}


\abstract{Cancer is a complex disease driven by dynamic regulatory shifts that cannot be fully captured by individual molecular profiling. We employ a data-driven approach to construct a coarse-grained dynamic network model based on hallmark interactions, integrating stochastic differential equations with gene regulatory network data to explore key macroscopic dynamic changes in tumorigenesis. Our analysis reveals that network topology undergoes significant reconfiguration before hallmark expression shifts, serving as an early indicator of malignancy. A pan-cancer examination across $15$ cancer types uncovers universal patterns, where \textit{Tissue Invasion and Metastasis} exhibits the most significant difference between normal and cancer states, while the differences in \textit{Reprogramming Energy Metabolism} are the least pronounced, consistent with the characteristic features of tumor biology. These findings reinforce the systemic nature of cancer evolution, highlighting the potential of network-based systems biology methods for understanding critical transitions in tumorigenesis.}




\maketitle

\section{Introduction}\label{sec:Introduction}

Cancer is a complex and dynamic disease characterized by unchecked cell proliferation, genomic instability, and the disruption of normal regulatory mechanisms \cite{Hanahan00,Hanahan11,Swanton24}. Recent advances have significantly refined our view of cancer as a systemic disease governed by intricate interactions among heterogeneous cellular populations, dysregulated molecular networks, and adaptive evolutionary processes \cite{Swanton24, Marusyk10, Ma'ayan17}. Conventional reductionist approaches centered on individual genetic alterations often fail to capture the emergent properties arising from the collective activity of interconnected functional modules \cite{Aguade-Gorgorio2024a}.

Complex systems theory offers a framework to address these challenges. Biological systems are composed of molecular networks and evolve through nonlinear interactions \cite{Bergen20, Strober19, Li11}, displaying both robustness to certain perturbations and emergent properties sensitive to collective perturbations, which cannot be interpreted via individual alterations \cite{Auyang98, Nicolis90, Alexander20}. Moreover, the low-dimensional hypothesis suggests that the behavior of a complex system can be represented by the so-called coarse-grained system preserving essential network dynamics and patterns, whose minimal component units are sets of individuals of the original system \cite{Tu21}. Additionally, although similar complex systems may display substantial differences at the microscopic level, they often exhibit analogous universality at the macroscopic scale \cite{Newman01092005,Bak1996,Kaneko2004b}. Therefore, exploring the similarities among different cancers from a complex systems perspective is valuable.

In the context of cancer as a complex system, the concept of hallmarks of cancer can provide a potentially reasonable coarse-graining methodology because the framework of hallmarks of cancer provides a foundational paradigm for understanding cancers from a macroscopic level \cite{Hanahan00,Hanahan11,Hanahan2022}. The concept of hallmarks of cancer was proposed by Hanahan and Weinberg in 2000 with six core features such as \textit{Self-Sufficiency in Growth Signals}, insensitivity to anti-growth signals, and others \cite{Hanahan00}, and later expanded to more hallmarks including \textit{Reprogramming Energy Metabolism}, \textit{Evading Immune Destruction}, \textit{Tumor-Promoting Inflammation}, and \textit{Genome Instability and Mutation} \cite{Hanahan11,Hanahan2022}. Since hallmarks are essentially functional descriptions, they can be represented as sets of genes with related functions within interaction networks whose dysregulation facilitates malignant progression \cite{Thibeault24,Swanton24,Jain2023}. Recent spatial transcriptomics studies further reveal that hallmark activities are spatially compartmentalized, forming interdependent ecological niches that drive tumor progression \cite{Sibai2025}, validating the importance of macroscopic view of cancer evolution, which is also supporting further model-driven study based on the architecture of hallmarks. Thus, the ``package'' of genes of hallmarks can systematically encapsulate cancer's phenotypic complexity through fundamental principles governing malignant transformation via sequential acquisition of functional capabilities \cite{Vegue23} and network-level interactions among these hallmarks during carcinogenesis \cite{Crosby22}.

Within this theoretical framework of complex systems of cancer \cite{Aguade-Gorgorio2024a,Kang2024}, this work implements a coarse-graining methodology, reducing intricate gene regulatory networks to hallmark-associated gene sets that collectively represent distinct oncogenic processes by mapping hallmarks to genes via GO terms and consequently constructing the regulatory networks of hallmarks through knowledge in the GRAND database \cite{Ben22}, thereby establishing the regulatory network of which hallmarks serving as network-level proxies to illuminate system-level shifts in tumorigenesis as the transition from homeostasis to malignancy. Then, we develop a macroscopic stochastic dynamic model to simulate the evolution of hallmark dynamics during tumorigenesis transitions from normal to malignant phenotypes across $15$ cancer types, based on which, computational methods such as Dynamic Network Biomarker (DNB) theory \cite{Chen12} and hierarchical clustering are employed to identify pan-cancer dynamic patterns. 

Our findings show that the activity of \textit{Tissue Invasion and Metastasis} exhibits the greatest difference between normal and tumor tissues, while \textit{Reprogramming Energy Metabolism} demonstrates a more conserved regulatory pattern during tumorigenesis. This finding is biologically meaningful and logical, as it is consistent with the role of these processes in cancer biology. Moreover, this phenomenon is observed across $15$ cancer types, indicating the universal patterns of tumorigenesis from the Hallmark perspective. In addition, the network structure changes, characterized by dynamic network biomarkers, emerge consistently prior to significant shifts in hallmark activity across all $15$ cancer types, suggesting that network reconfiguration occurs earlier than changes in absolute gene expression levels. These insights enhance our understanding of cancer as a complex adaptive system and provide a framework for anticipating critical transitions in tumor progression.

\section{Results}\label{sec:Results}

\subsection{The hallmark network and the mathematical model for its evolution dynamics}
\begin{figure}[pthb]
    \centering
    \includegraphics[width=1\textwidth]{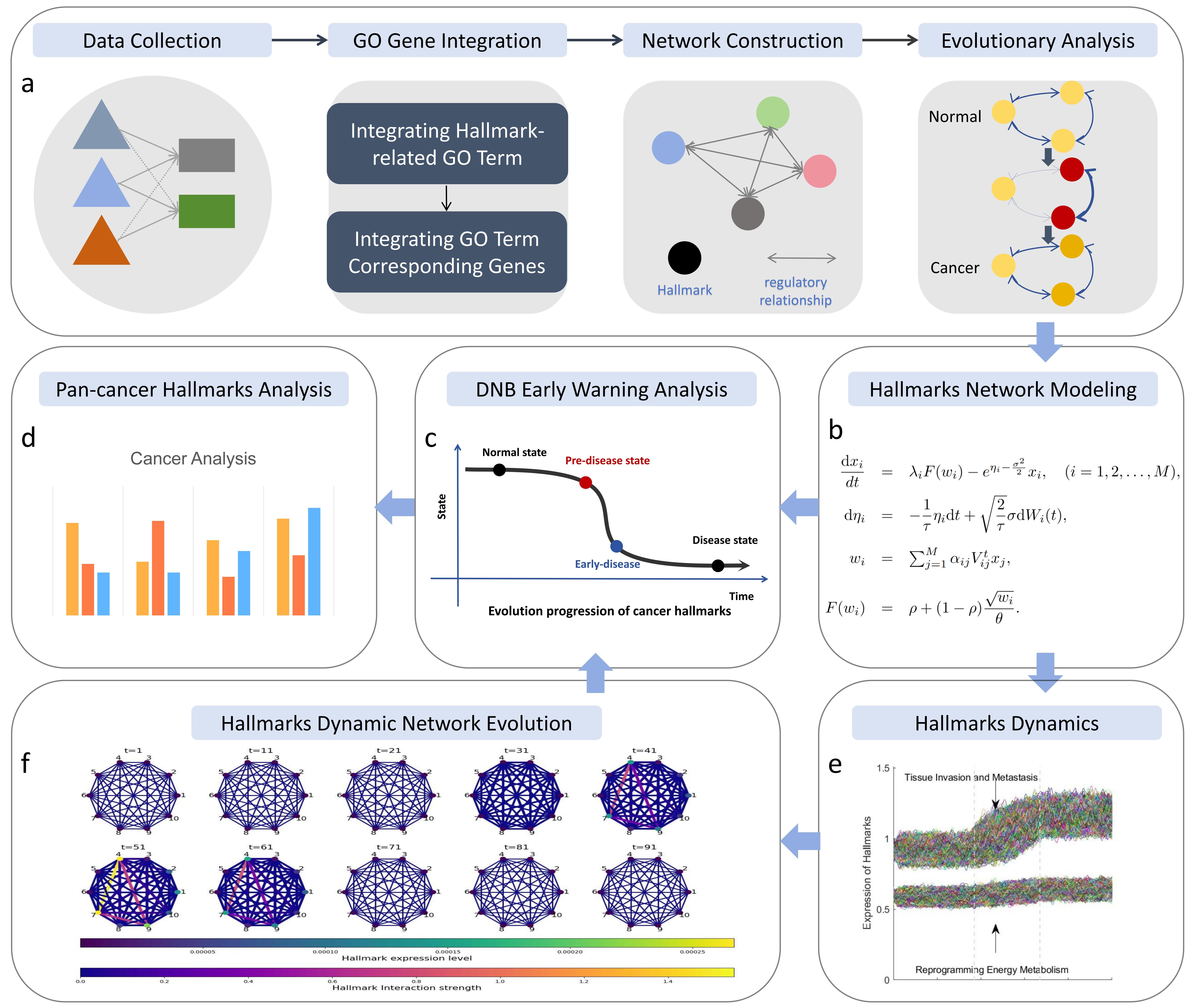}
    \caption{
      \textit{System-level modeling of hallmark network evolution.}  
      \textbf{(a)} The framework of this work. Upon integrating gene expression data with their regulatory interactions corresponding, the interaction network of hallmarks is constructed by coarse-graining of gene regulatory networks. Dynamic equations are established to simulate the evolution of hallmark network from normal to cancerous states. Then, Pan-cancer analysis across $15$ cancer types is conducted to explore commonalities and differences in the evolutionary trajectories during tumorigenesis. 
      \textbf{(b)} Dynamical equations for simulating state transitions between normal and cancerous phenotypes. 
      \textbf{(c)} Tipping point detection using DNB theory at critical transition points. 
      \textbf{(d)} Pan-cancer analysis of evolutionary trajectories across $15$ cancer types. 
      \textbf{(e)} Different dynamic patterns of hallmarks' evolutionary trajectories from normal to cancerous states. 
      \textbf{(f)} Temporal evolution of Hallmark strength and time-varying network of their interactions. 
    }
    \label{fig:framework}
\end{figure}

Cancer progression emerges from the dysregulation of interconnected molecular networks, necessitating a systemic perspective to unravel its evolutionary dynamics. By coarse-graining the gene regulatory network based on the hallmarks of cancer, a low-dimensional macroscopic dynamic model is established (Figure \ref{fig:framework}) to capture the transition from normal to malignant states. Inspired by the low-rank hypothesis of complex systems \cite{Thibeault24}, it is reasonable to represent cancer cells by a network of hallmarks, which is defined as sets of genes curated from GO terms \cite{Ashburner2000a,Aleksander2023,Carbon09}, while the interactions between the hallmarks are extracted from the gene regulation network datasets, such as the GRAND database \cite{Ben22} which has data of both tumor tissues and corresponding normal tissues.

To construct the interactions between hallmarks from the gene regulatory networks, the hallmark-associated GO terms are identified through previous studies \cite{Plaisier12} where a well-established mapping between GO-terms and hallmarks is verified, and consequently, the gene sets corresponding to those GO-terms are used as the hallmarks. Secondly, given the quantitative interaction between the individual genes from the GRAND database \cite{Ben22}, the interaction strengths between two gene sets (i.e., hallmarks) are given by aggregating the interaction between each pair of genes from the two gene sets.

Finally, to simulate hallmark dynamics based on the interaction network of hallmarks, a set of stochastic differential equations incorporating Ornstein-Uhlenbeck noise was established through a general framework of modeling the noising gene regulatory network dynamics \cite{Liu2021q} as shown in Equations \ref{eq:model} and Figure \ref{fig:framework}b.
\begin{equation}
\begin{array}{rcl}
    \dfrac{\mathrm{d} x_i}{dt} &=& \lambda_i F(w_i) - e^{\eta_i - \frac{\sigma^2}{2}} x_i, \quad (i=1,2, \cdots, M), \\[0.3cm]
    \mathrm{d} \eta_i &=& -\dfrac{1}{\tau} \eta_i \mathrm{d} t + \sqrt{\dfrac{2}{\tau}} \sigma \mathrm{d} W_i(t), \\[0.5cm]
    w_i &=& \sum_{j=1}^{M} \alpha_{ij} V^t_{ij} x_j, \\[0.5cm]
    F(w_i) &=& \rho + (1 - \rho) \dfrac{\sqrt{w_i}}{\theta}. \\[0.5cm]
\end{array}
\label{eq:model}
\end{equation}
where $V^t_{ij}$ represents time-dependent regulatory strengths, interpolated between normal and cancer states extracted from GRAND and $w_i$ represents the expression quantification of the $i$-th hallmark. More details are described in the method section and supplementary materials. The changing $V^t_{ij}$ enables us to quantify three distinct phases in carcinogenesis: an initial stationary phase mimics the healthy homeostatic states, followed by a critical transition marked by network reconfiguration and final cancer (abnormal) states (Figure \ref{fig:framework}e). 

Taking the gastric adenocarcinoma data in the GRAND database as an example, 10,000 trajectories of hallmark network evolution are obtained by stochastic simulation (Figure \ref{fig:Hallmarks expression}a). All the averages of the hallmark levels are higher in cancerous states than those in normal states (Figure \ref{fig:Hallmarks expression}c), which is consistent with the conceptual framework of the concept of hallmarks.  Moreover, to simulate a single patient with heterogeneous cell populations, 1,000 randomly selected trajectories are aggregated to represent an individual composed of heterogeneous cells because the dynamics of hallmark network mimics the expression patterns of gene sets in individual cells with similar states. This setting enables the construction of dynamic expression profiles.

\subsection{Differential dynamics of hallmarks during tumorigenesis}
\begin{figure}[tbhp]
    \centering
    \subfigure[]{\includegraphics[width=13.5cm]{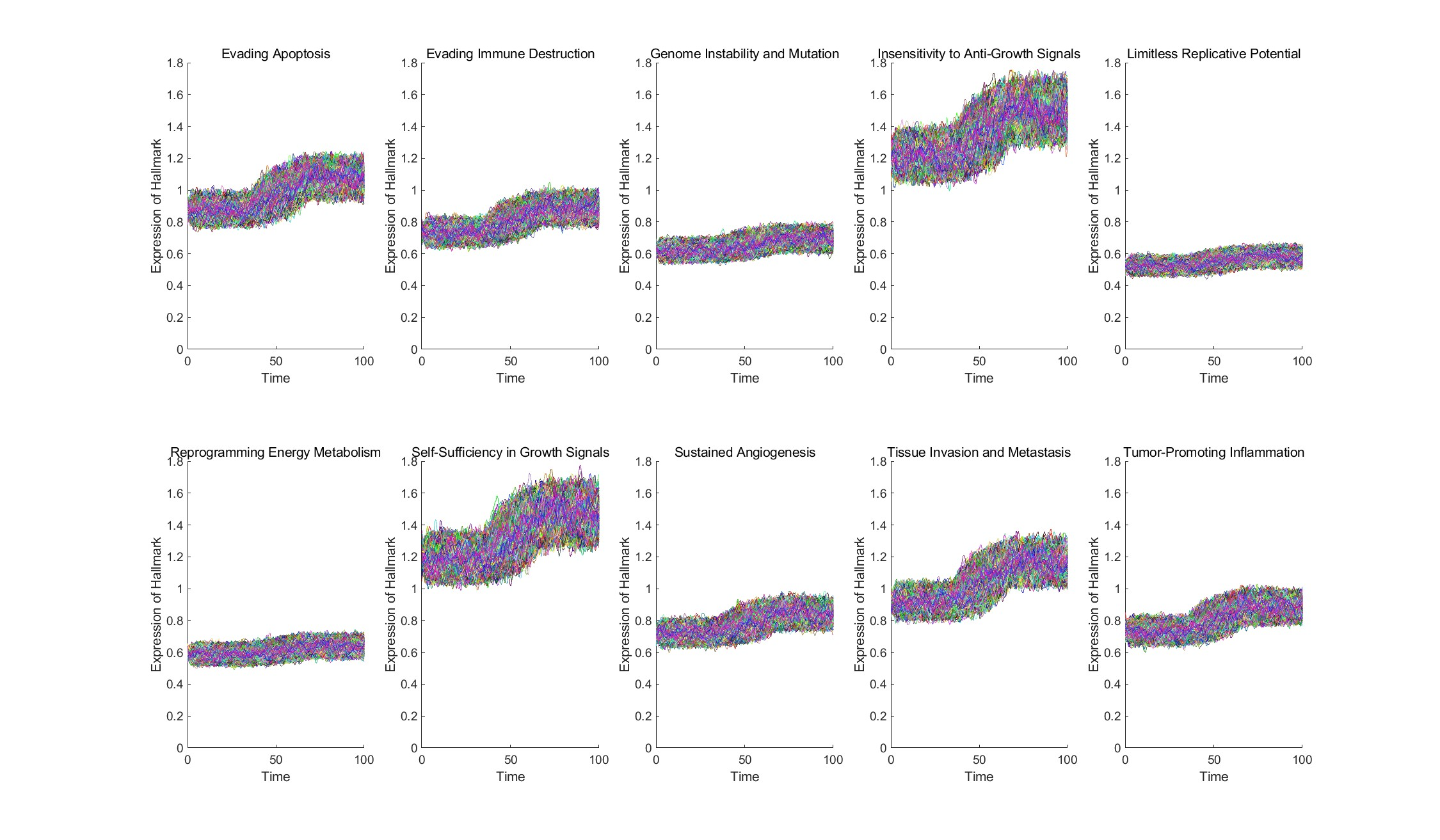}}\\
    \subfigure[]{\includegraphics[width=13.5cm]{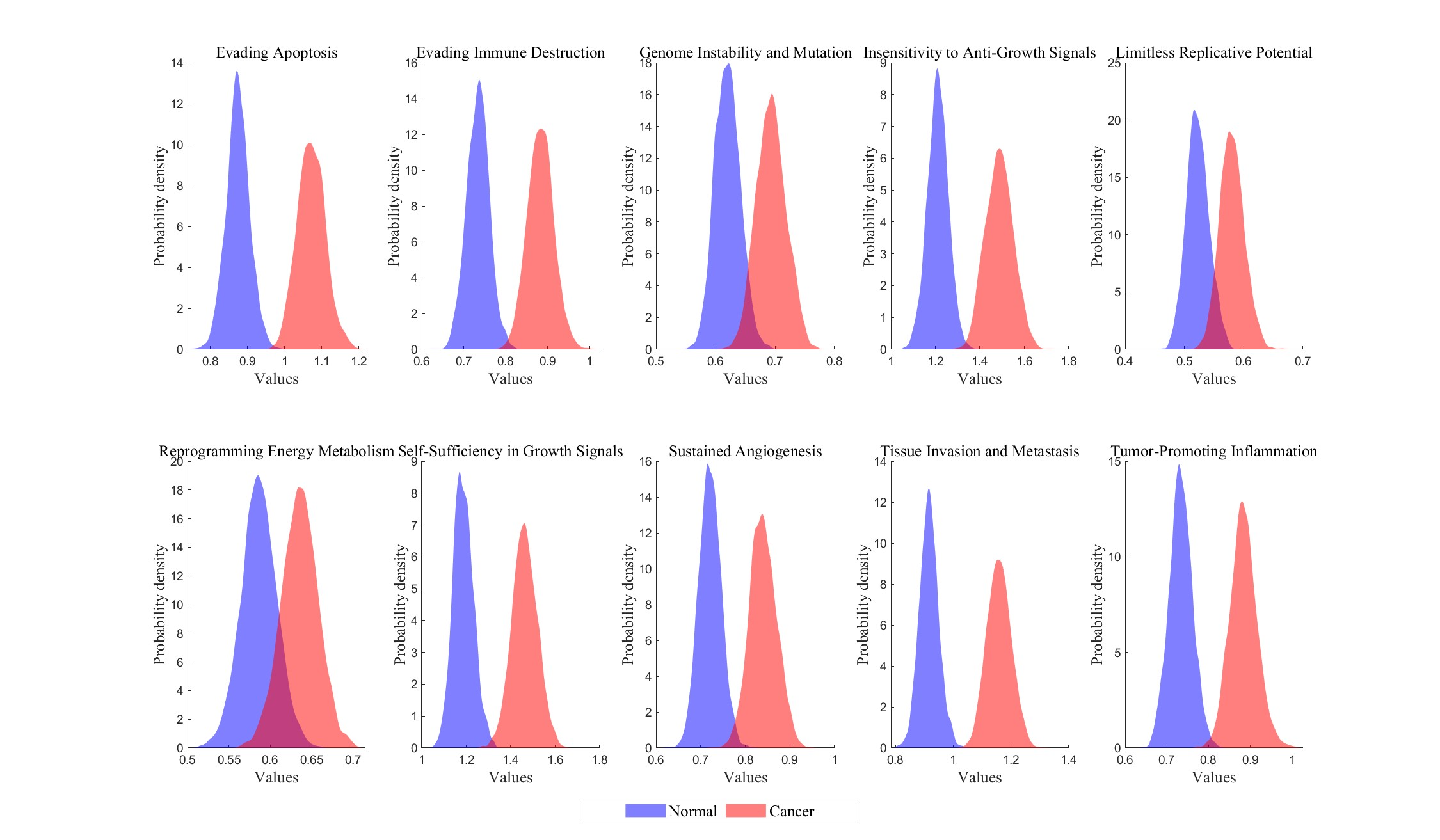}}
    \label{fig:Hallmarks expression}
    \subfigure[]{\includegraphics[width=6.3cm]{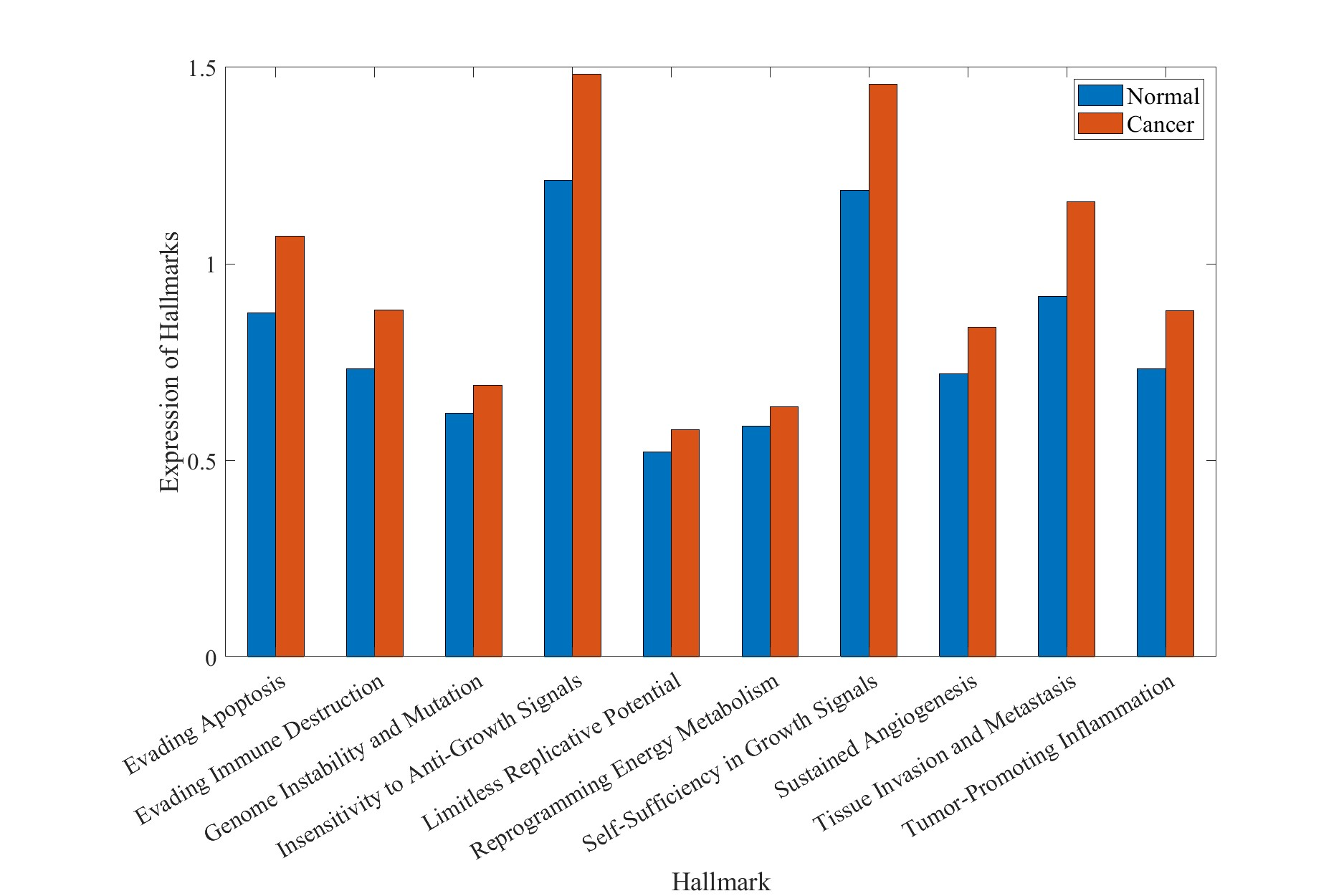}}
    \subfigure[]{\includegraphics[width=6.3cm]{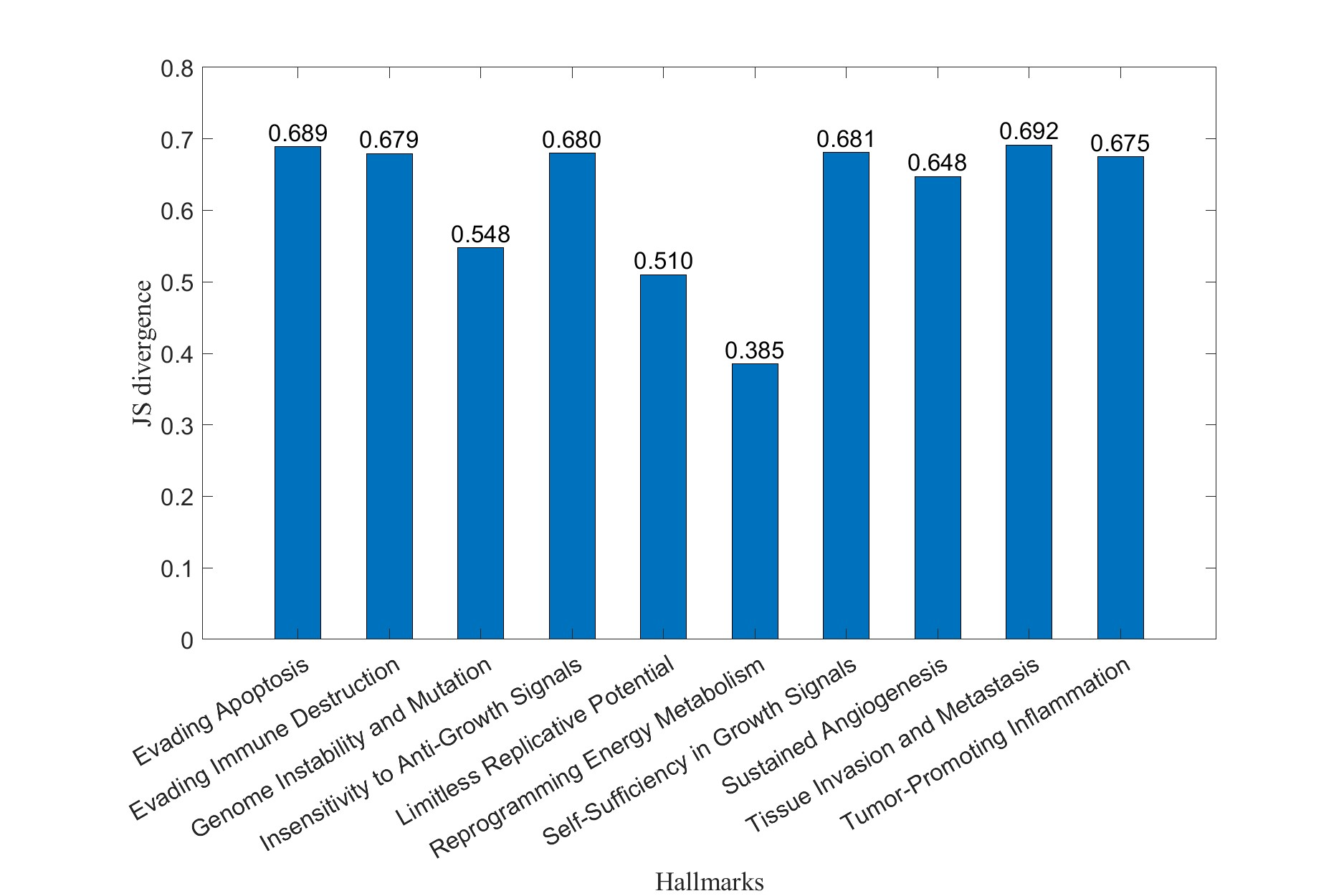}}
    \caption{Hallmark dynamics of gastric adenocarcinoma tumorigenesis.
     \textbf{(a)} Trajectories of the hallmark levels from normal to cancerous states. Each subfigure contains 10,000 simulations.
     \textbf{(b)} The distributions of hallmarks in normal and cancerous states, i.e., initial states and final states in the simulations, respectively. 
     \textbf{(c)} The average levels of the hallmarks in normal and cancerous states.
     \textbf{(d)} The Jensen–Shannon divergences (JS divergences) of the ten hallmarks between normal and cancerous states.}
\end{figure}

According to Figure \ref{fig:Hallmarks expression}a, the dynamic evolution of hallmarks during tumorigenesis of gastric adenocarcinoma shows distinct patterns of regulatory divergence between normal and malignant states. To quantify hallmark-specific divergences, the distributions of hallmark expression levels at healthy and cancerous states are examined (Figure \ref{fig:Hallmarks expression}b) and compared using the Jensen--Shannon (JS) divergence (Figure \ref{fig:Hallmarks expression}d). 

Accordingly, \textit{Tissue Invasion and Metastasis} exhibits the most significant difference between normal and cancerous states, whereas \textit{Reprogramming Energy Metabolism} shows only minimal differences (Figure \ref{fig:Hallmarks expression}a,b,d). The heterogeneity in hallmark dynamics aligns with their distinct contributions to tumorigenesis in a manner that reflects both shared and cancer-specific mechanisms \cite{Hanahan00,Hanahan11}.

Specifically, \textit{Tissue Invasion and Metastasis} demonstrates the greatest separation between normal and cancer groups. This hallmark is linked to key processes such as wound healing, negative regulation of cell adhesion, epithelial-to-mesenchymal transition (EMT), and cell migration \cite{Chaffer16,Brabletz01,Bourboulia10}. In normal tissues, cell adhesion and extracellular matrix integrity uphold tissue structure, preventing aberrant cell dissemination. However, cancer cells must overcome these constraints by downregulating epithelial adhesion molecules (e.g., E-cadherin) while upregulating mesenchymal markers (e.g., N-cadherin, vimentin), a hallmark of EMT that enhances motility and invasiveness \cite{Chaffer16, Brabletz01}. Furthermore, cancer cells hijack tissue remodeling programs, leading to sustained ECM degradation and integrin-mediated signaling, thereby facilitating invasion and metastasis \cite{Bourboulia10, villalobo2020role}. These pathways have been shown to equip malignant cells with invasive capacities, which are adaptations unique for cancer cells to breach tissue barriers promoting distant metastasis \cite{villalobo2020role,kleiner1999matrix}, collectively account for the significant expression divergence observed in \textit{Tissue Invasion and Metastasis}. 

In contrast, hallmarks such as \textit{Reprogramming Energy Metabolism} display smaller expression differences. Although \textit{Reprogramming Energy Metabolism} constitutes a central hallmark of cancer, fundamental metabolic adaptations such as glycolysis, often associated with the Warburg effect, are also activated in normal cells under hypoxia or stressed conditions \cite{Vander09,DeBerardinis20,Sun19}. In normal cells, for example, hypoxia-inducible factors (HIFs) orchestrate metabolic adaptations by upregulating glycolysis, angiogenesis, and glutamine metabolism to maintain energy homeostasis \cite{Lee20,Kierans21,Eales16}. 
Proliferating embryonic and immune cells similarly enhance glycolysis to support heightened biosynthetic demands \cite{Eales16}. The conservation of the metabolic mechanisms reduces the difference between normal and malignant states, leading to smaller expression differences in \textit{Reprogramming Energy Metabolism} compared to hallmarks that drive invasion and structural transformation. These conserved mechanisms result in substantial overlap between metabolic flexibility in normal cells and pathological reprogramming in cancer \cite{Eales16,DeBerardinis20}.  Therefore, being an inherent feature of proliferating cells rather than a unique property of cancer, the shared regulatory mechanism of metabolic plasticity explains why the \textit{Reprogramming Energy Metabolism} hallmark exhibits smaller divergence between normal and malignant states compared to invasion/metastasis-related hallmarks.

Additional hallmarks, including \textit{Evading Apoptosis} and \textit{Self-Sufficiency in Growth Signals}, also exhibit notable changes. Evading cell death often involves suppression of pro-apoptotic signals (e.g., p53) and overactivation of anti-apoptotic genes (e.g., BCL-2 family) \cite{pistritto2016apoptosis}, enabling cancer cells to survive under conditions that would normally trigger apoptosis. Similarly, the heightened divergence in \textit{Self-Sufficiency in Growth Signals} highlights how persistent activation of growth factor pathways, including EGFR, can circumvent homeostatic constraints on cell proliferation \cite{wee2017epidermal}. In contrast, \textit{Limitless Replicative Potential} and \textit{Genome Instability and Mutation} show smaller differences at the gene expression level in early or mid-stages of tumorigenesis, potentially due to partial overlap with normal proliferative mechanisms or later-stage emergence \cite{albanell1997high,shay2016role}.

So it is evident that tumor progression relies on both early disruptors of cell survival and proliferation (e.g., \textit{Evading Apoptosis}, \textit{Self-Sufficiency in Growth Signals}), as well as on hallmark traits that underpin advanced dissemination (e.g., \textit{Tissue Invasion and Metastasis}). Meanwhile, those hallmarks that exhibit smaller distributional shifts (such as \textit{Reprogramming Energy Metabolism}) may still play an essential role but follow regulatory pathways shared with certain normal proliferative processes.

Notably, as the computational model is built upon the real-world data inferred GRAND datasets \cite{Ben22}, the simulated results in this study are consequences of data-driven observations. Therefore, the above analysis reveals that, from the macroscopic view, in the tumorigenesis of gastric adenocarcinoma, \textit{Tissue Invasion and Metastasis} is central to malignant progression, whereas certain metabolic and replicative processes partially overlap with normal cell physiology. Such nuanced insights into hallmark-specific dynamics foster a better understanding of how cancers emerge and evolve.

\subsection{Early Network Reconfiguration Precedes Hallmark Expression Changes}
\begin{figure}[tbhp]
    \centering
      \subfigure[]{\includegraphics[width=0.9\textwidth]{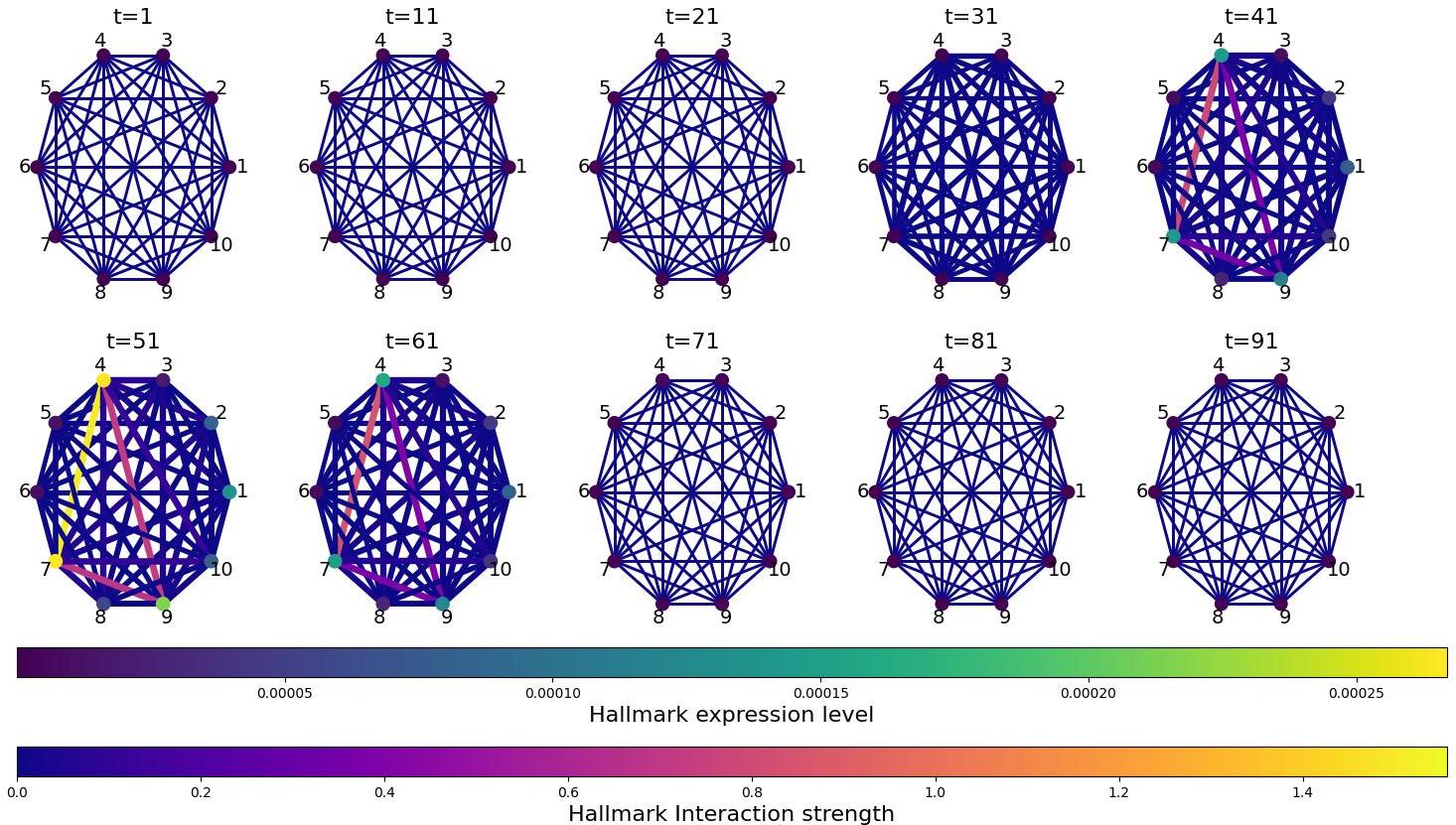}}

      \subfigure[]{\includegraphics[width=0.6\textwidth]{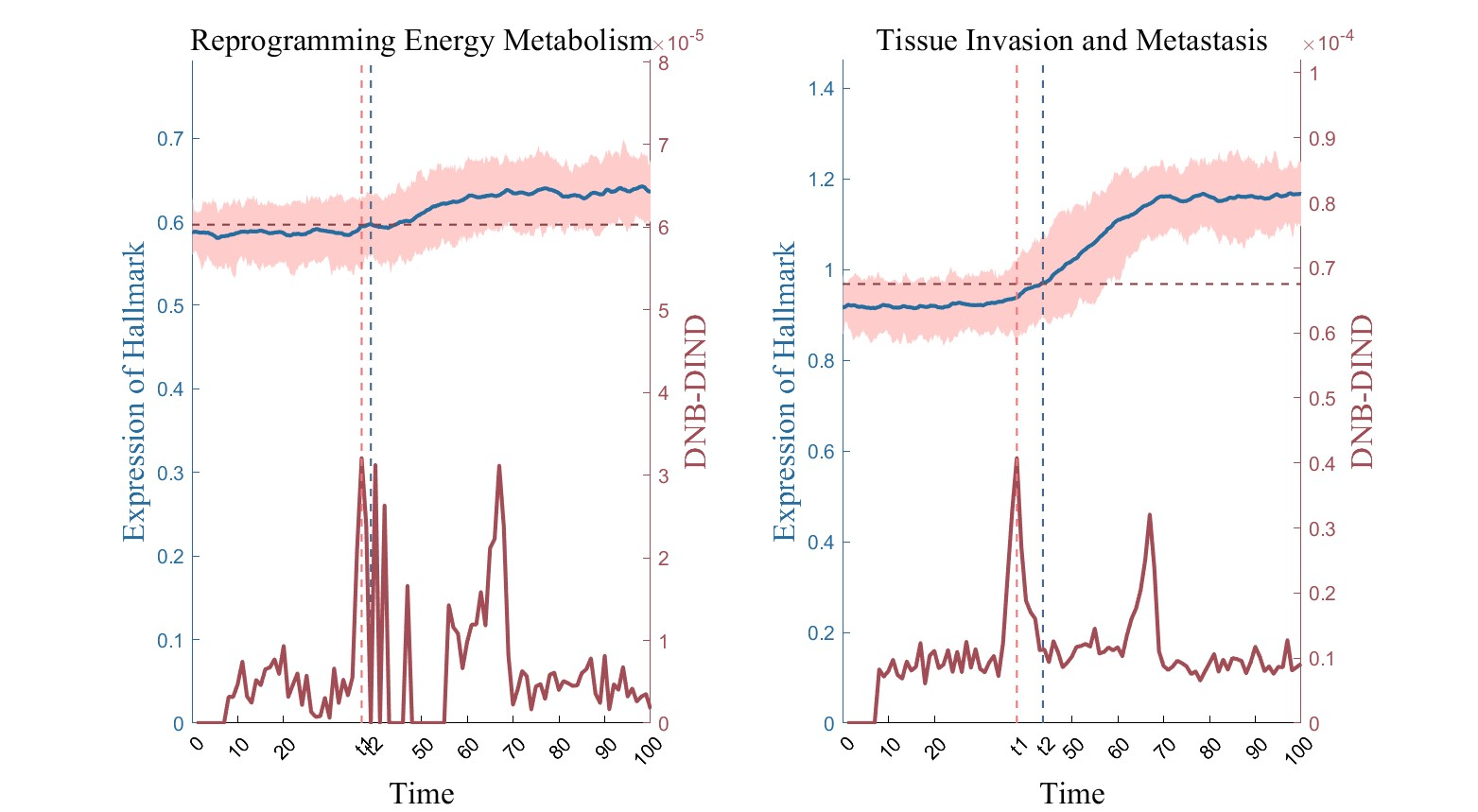}}
      \subfigure[]{\includegraphics[width=0.35\textwidth]{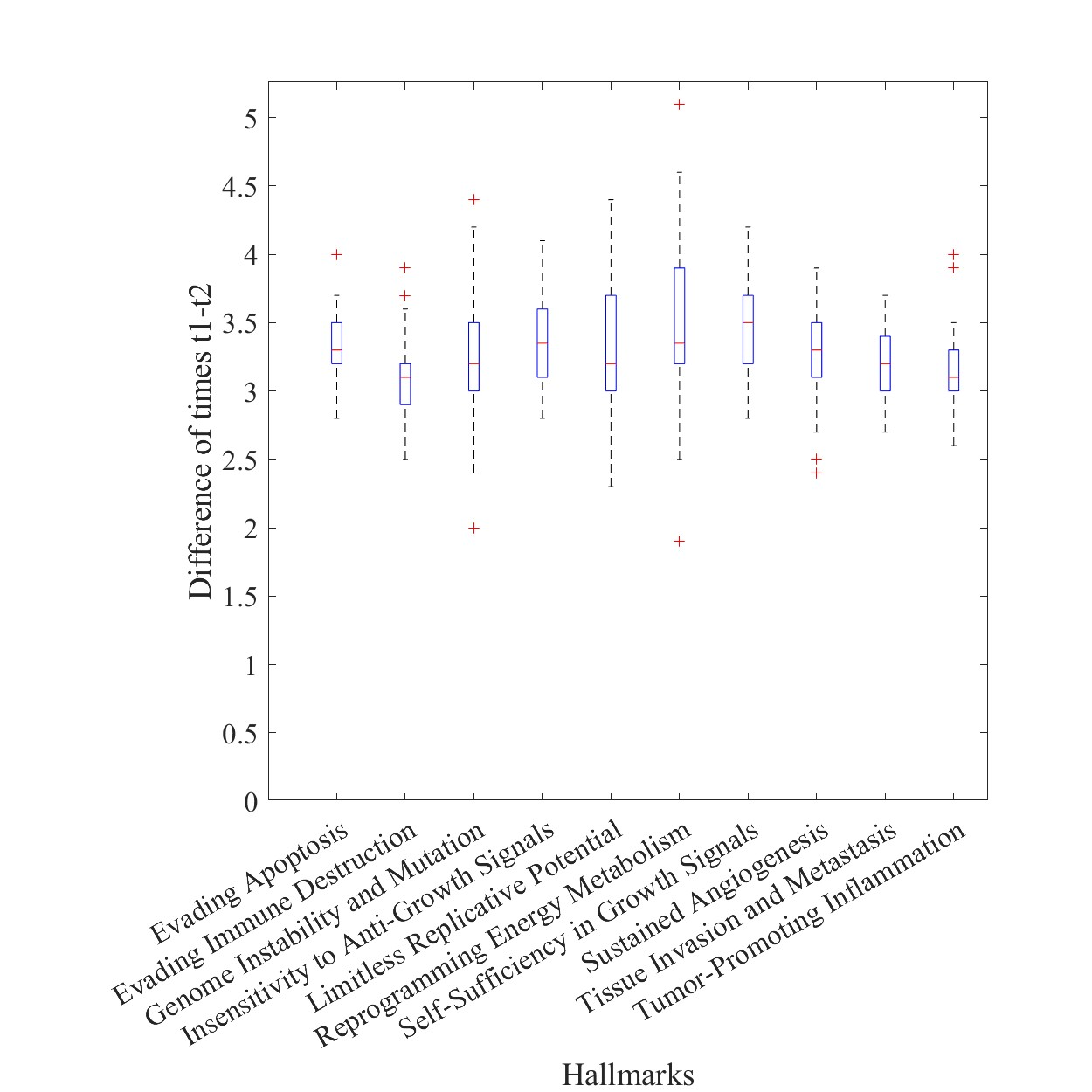}}
    \caption{
       \textbf{(a)} Hallmark networks during the transition from normal to cancerous states. The expression level of hallmarks and the strength of interactions are indicated by colorbar at the bottom of the plots. \textbf{(b)} Comparison between the averaged trajectories and the DNB score (DIND). The blue curve represents the average expression levels of the hallmarks, while the red curve depicts the DNB-derived score that quantifies changes in the network structure. Here, time point $t_1$ marks the first peak in the DIND score—interpreted as the early critical point of cancer progression—whereas $t_2$ denotes the moment when a hallmark’s expression level exceeds the critical threshold of the normal state, defined as 1.2 times the normal steady-state level. \textbf{(c)} Distribution of the time differences ($t_2-t_1$) between the initial critical point and the point at which a hallmark's expression level surpasses the normal-state threshold.
       }
       \label{fig:network evolution}
\end{figure}

Cancer is a complex disease that can be conceptualized as a dynamic system, with key transitions manifesting as state shifts at bifurcation (critical) points \cite{Aguade-Gorgorio2024a}. However, the levels of individual molecules or biomarkers may not exhibit significant changes near these critical points. In contrast, the interaction network among these molecules or biomarkers can undergo dramatic structural reconfigurations \cite{Chen12}. Therefore, the simulation results in this study, the alterations in the network structure of the Hallmarks are also investigated. To this end, the DNB theory is employed to characterize the temporal changes in the network structure \cite{Chen12}. DNB theory posits that as a biological system approaches a critical transition, a subset of key molecules begins to exhibit significantly enhanced fluctuations and increased correlations, forming a dynamically coherent module,  which serves as a sensitive early-warning signal of an imminent critical state \cite{Chen12,Peng22,Liu2022i,Zhang2024d,Kang2024}. Among many DNB indices, as the direct interaction network-based divergence (DIND) method \cite{Peng22} captures topological reorganization, DIND is used for revealing the network rewiring initiation during the pre-disease phase (Figure \ref{fig:network evolution}).

As shown in Figure \ref{fig:network evolution}b, for the most significant differential expression hallmark, \textit{Tissue Invasion and Metastasis}, the time point of reorganization of hallmark network ($t_1$) occurs approximately three time units prior to that of the average expression levels crossing the normal threshold ($t_2$). Furthermore, although the \textit{Reprogramming Energy Metabolism} hallmark exhibits the smallest difference (Figure \ref{fig:Hallmarks expression}), the DIND score still occurs earlier than that of the expression level crossing the threshold, indicating that the structure of the hallmark network, which reflects the complex inter-relationships among functional modules, is a more sensitive early indicator of the impending malignant transition than the changes in the quantitative expression of each hallmark.

The early remodeling of network interactions supports the hypothesis that changes in regulatory connectivity serve as a precursor to overt phenotypic shifts in tumorigenesis. Such early network reconfiguration potentially provides valuable novel therapeutic strategies, such as targeting regulatory hubs before malignant phenotypes emerge, for early intervention in the processes of tumorigenesis.


\subsection{Pan-cancer Hallmark Network Dynamics Reveal Sequential Progression Patterns}
To investigate the similarities and differences in hallmark dynamics across diverse cancer types, we analyze the data of $15$ cancer types from the GRAND database including Stomach Adenocarcinoma (STAD), Kidney Renal Papillary Cell Carcinoma (KIRP), Kidney Renal Clear Cell Carcinoma (KIRC), Lung Adenocarcinoma (LUAD), Lung Squamous Cell Carcinoma (LUSC), Pheochromocytoma and Paraganglioma (PCPG), Cutaneous Melanoma (SKCM), Thyroid Cancer (THCA), Uveal Melanoma (UVM), Acute Myeloid Leukemia (LAML), Adrenocortical Carcinoma (ACC), Low-Grade Glioma (LGG), Esophageal Cancer (ESCA), Head and Neck Cancer (HNSC), and Kidney Chromophobe (KICH). 

\begin{figure}[H]
    \centering
    \subfigure[]{\includegraphics[width=0.49\textwidth]{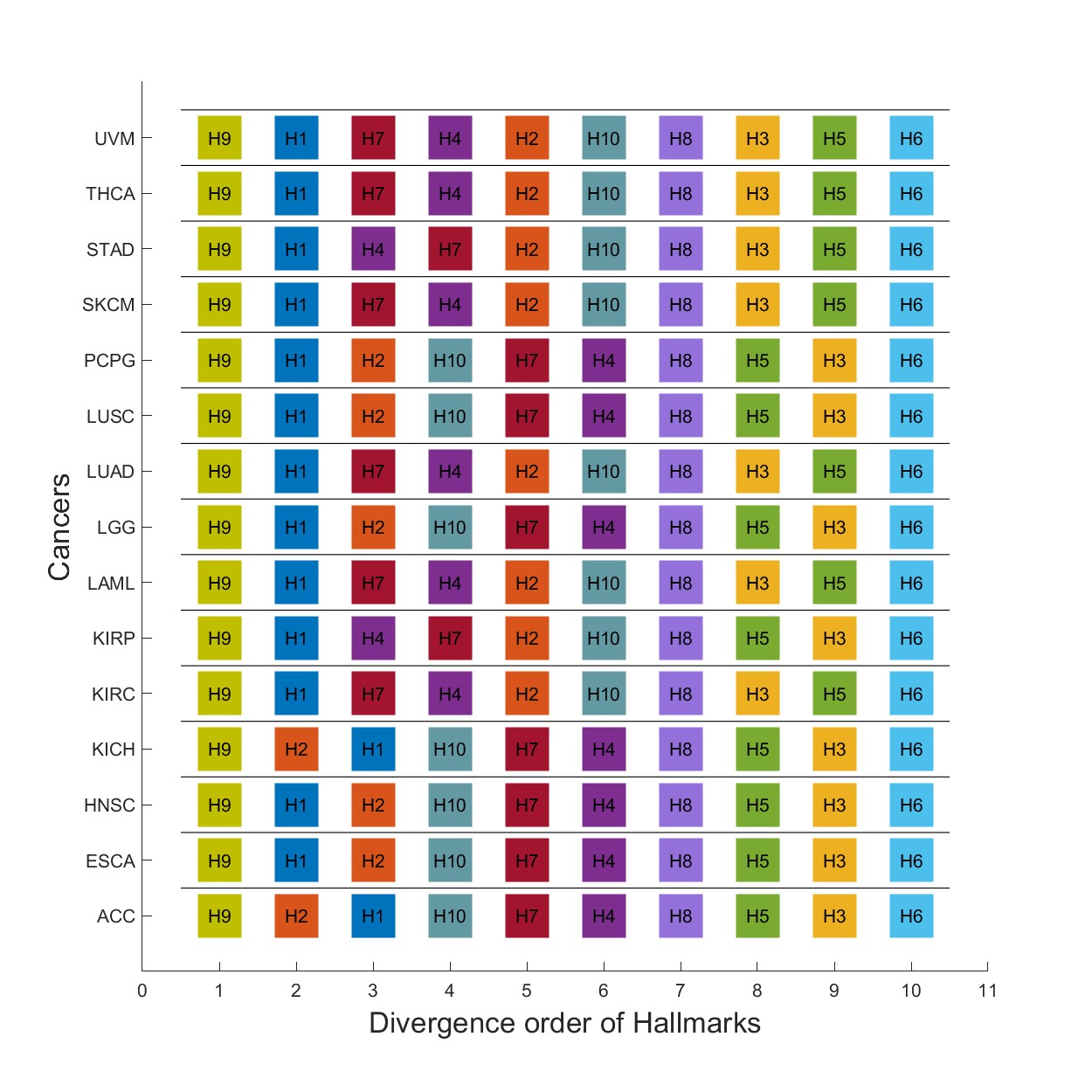}}
    \subfigure[]{\includegraphics[width=0.49\textwidth]{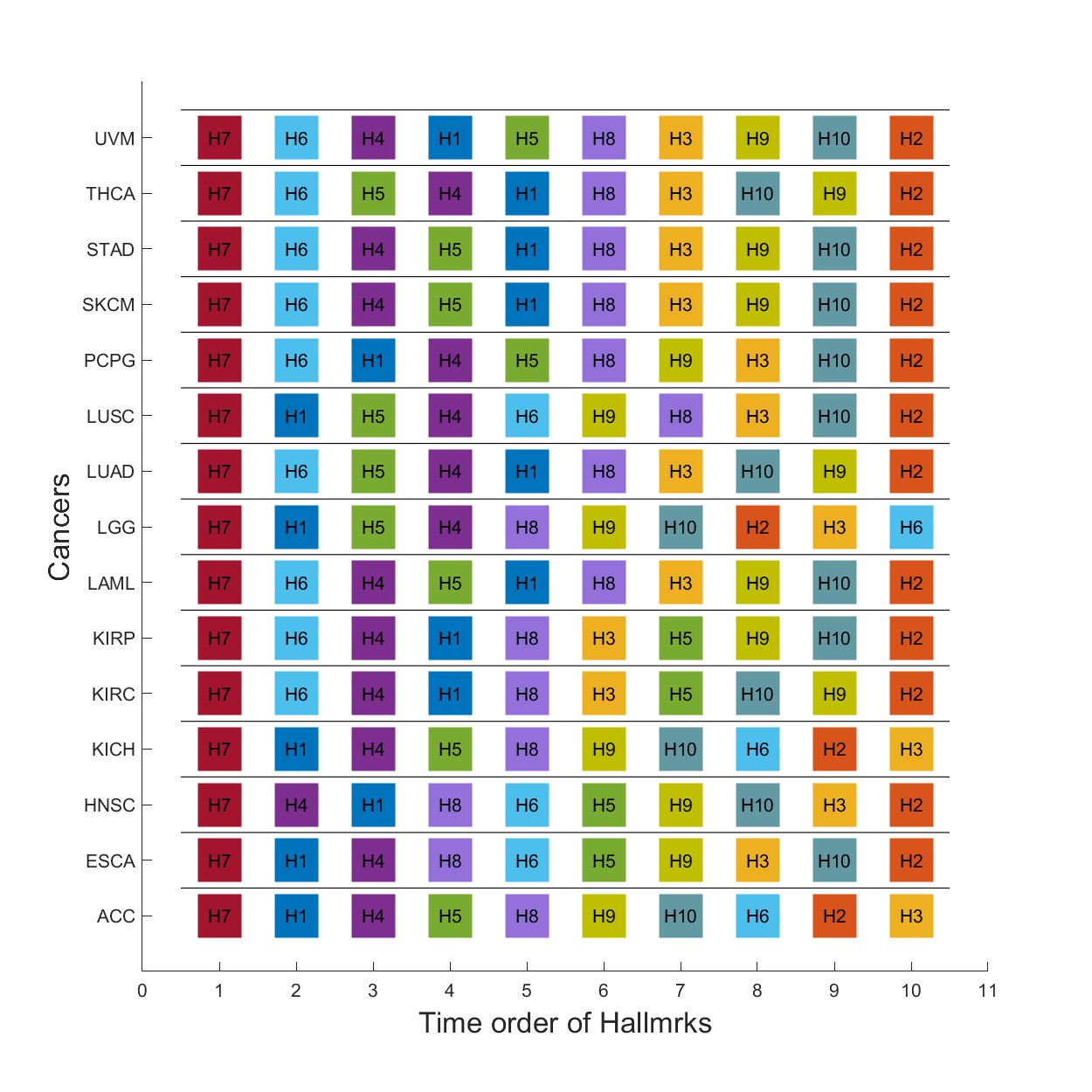}}\\
    \subfigure[]{\includegraphics[width=0.51\textwidth]{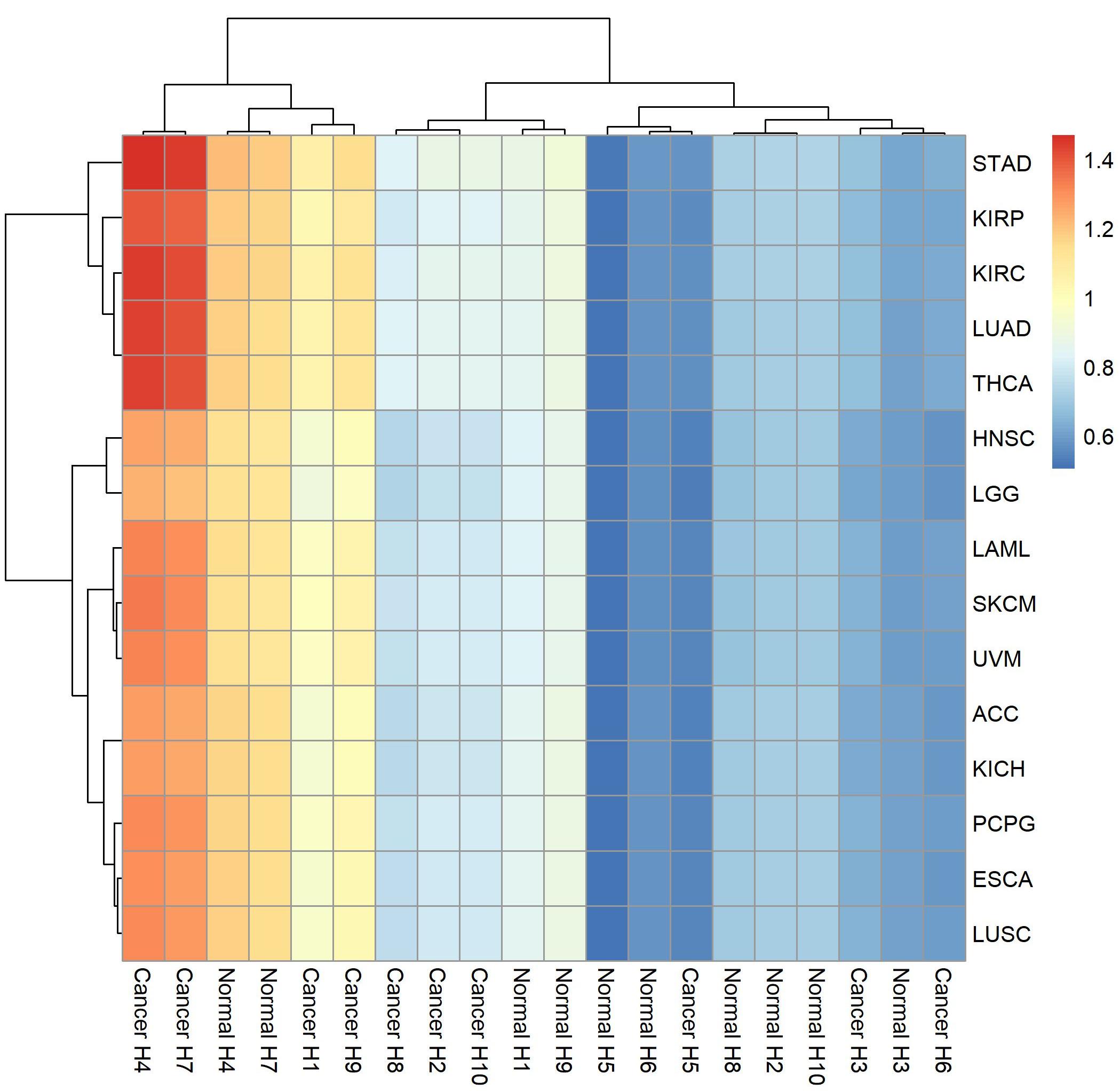}}
    \subtable[]{
    \footnotesize
    \begin{tabular}[b]{|c|m{0.33\textwidth}|}
        \hline
        H1	&	\textit{Evading Apoptosis}	\\\hline
        H2	&	\textit{Evading Immune Destruction}	\\\hline
        H3	&	\textit{Genome Instability and Mutation}	\\\hline
        H4	&	\textit{Insensitivity to Anti-Growth Signals}	\\\hline
        H5	&	\textit{Limitless Replicative Potential}	\\\hline
        H6	&	\textit{Reprogramming Energy Metabolism}	\\\hline
        H7	&	\textit{Self-Sufficiency in Growth Signals}	\\\hline
        H8	&	\textit{Sustained Angiogenesis}	\\\hline
        H9	&	\textit{Tissue Invasion and Metastasis}	\\\hline
        H10	&	\textit{Tumor-Promoting Inflammation}	\\\hline
    \end{tabular}
    }\\
    \caption{
      Analysis of commonalities and temporal patterns across $15$ cancer types. 
      \textbf{(a)} The expression distributions between normal and cancer states for the ten hallmarks are ranked by the JS divergences. Here the notations of hallmarks (H1-H10) are named as the order of appearance in Figure \ref{fig:Hallmarks expression}. 
      \textbf{(b)} Hallmarks are ranked by the advance in network reconfiguration (early-warning) time relative to the point when their expression levels exceed the normal range. The notations of hallmarks are similar to those in (a).
      \textbf{(c)} Hierarchical clustering of hallmark expression levels in normal and cancerous states. 
      \textbf{(d)} The hallmarks corresponding to notations (H1-H10).
    }
    \label{fig:Halmarksorder}
\end{figure}

For each cancer type, the differential expression of hallmarks between normal and cancerous states is quantified using JS divergence (Figure \ref{fig:Halmarksorder}a). The results show that \textit{Tissue Invasion and Metastasis} consistently exhibits the largest difference, followed by \textit{Evading Apoptosis} and \textit{Self-Sufficiency in Growth Signals}. These marked changes indicate that these hallmarks play critical roles in cancer progression, whereas \textit{Reprogramming Energy Metabolism}, \textit{Limitless Replicative Potential}, and \textit{Genome Instability and Mutation} display smaller differences. Notably, the \textit{Tissue Invasion and Metastasis} (H9) and \textit{Reprogramming Energy Metabolism} (H6) hallmarks represent the most and least pronounced differences between normal and cancerous states, respectively, across all cancer types examined in this study. This observation suggests that despite the divergent genetic differences among various cancers \cite{Roehrig2024,Zhu2023a,Kandoth2013a,Tan2015a,Martinez2015a}, common dynamic patterns are present, which underscores the significant and unified role of considering the Hallmarks (i.e., gene sets) as an integrated whole in tumorigenesis and tumor evolution. Consequently, future research should move beyond merely analyzing differential expression at the level of individual genes between normal and tumor tissues, and instead explore the collective impact of synergistically functioning gene groups on tumor evolution. In other words, investigating strategies to target groups of genes with similar functions may hold substantial potential for advancing future cancer therapies.

The conserved ordering of hallmark activation across cancer types suggests evolutionary constraints on tumor progression pathways, where hallmarks provide a broad perspective for identifying common patterns across cancer types. The substantial differential expression in \textit{Tissue Invasion and Metastasis} indicates its central role in cancer cell invasion and metastasis. Key biological processes associated with this hallmark include extracellular matrix degradation mediated by matrix metalloproteinases (MMPs), alterations in cell adhesion regulated by integrins and cadherins, and cytoskeletal remodeling \cite{villalobo2020role, kleiner1999matrix}. These adaptations enable cancer cells to breach the basement membrane and invade surrounding tissues. Similarly, the significant alteration in \textit{Evading Apoptosis} indicates that cancer cells often suppress pro-apoptotic pathways (e.g., via p53 signaling) while upregulating anti-apoptotic genes (e.g., the BCL-2 family) \cite{pistritto2016apoptosis}, thus enhancing their survival. The pronounced difference in \textit{Self-Sufficiency in Growth Signals} reflects the cancer cells’ ability to bypass normal growth control through aberrant activation of growth factor signaling (such as EGFR) \cite{wee2017epidermal}. In contrast, the relatively smaller differences in \textit{Reprogramming Energy Metabolism}, \textit{Limitless Replicative Potential}, and \textit{Genome Instability and Mutation} may indicate that these processes are shared with other rapidly proliferating cells— for instance, the Warburg effect is also observed in embryonic and immune cells \cite{Sun19}, and telomerase activation is typically a late-stage event \cite{albanell1997high, shay2016role}.

In addition, the time difference between the onset of network reconfiguration, assessed by DIND score, and the time point when hallmark expression levels exceed the normal threshold is also investigated (Figure \ref{fig:Halmarksorder}b). In most cancers, the time point of the network reorganization is observed to precede the overt expression changes in hallmarks such as \textit{Self-Sufficiency in Growth Signals}, \textit{Reprogramming Energy Metabolism}, and (where applicable) \textit{Insensitivity to Antigrowth Signals}. This temporal precedence indicates that the alteration of inter-hallmark regulatory interactions occurs before the quantitative changes in individual hallmark activities, thereby serving as a sensitive early indicator of malignant transition.

To further quantify the minor distinction in patterns in hallmark dynamics among cancer types, hierarchical clustering analysis is applied (Figure \ref{fig:Halmarksorder}c). STAD, KIRP, KIRC, LUAD, and THCA cluster together, exhibiting insensitivity to antigrowth signals and self-sufficiency in growth, often via \textit{TP53} and \textit{RB1} inactivation and TGF-$\beta$ suppression. They sustain proliferation through autocrine/paracrine growth factors (e.g., EGF, FGF) and receptor alterations (EGFR, HER2), activating PI3K/AKT/mTOR and MAPK/ERK pathways. In contrast, LUAD frequently harbors EGFR/KRAS mutations affecting these pathways, while LUSC, driven by smoking-related \textit{CDKN2A} mutations, disrupts cell cycle regulation and apoptosis \cite{Zengin21}. Moreover, \textit{CEP55} co-expresses with cell cycle and DNA replication genes in LUAD but not in LUSC \cite{Fu20}, underscoring differences in tissue origin, cell type, and microenvironmental adaptation.

\begin{figure}[pthb]
    \centering
    \subfigure[]{\includegraphics[width=0.39\textwidth]{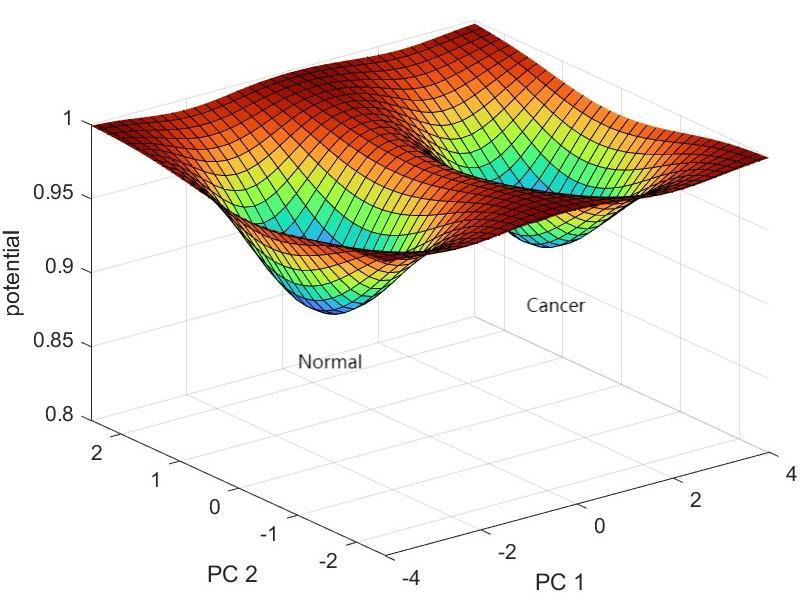}}
    \subfigure[]{\includegraphics[width=0.39\textwidth]{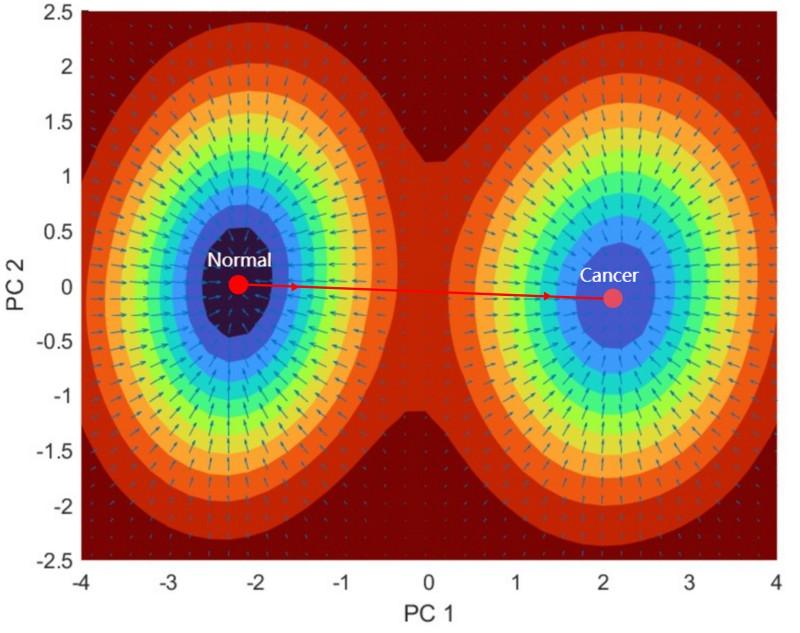}}\\
    \subfigure[]{\includegraphics[width=0.78\textwidth]{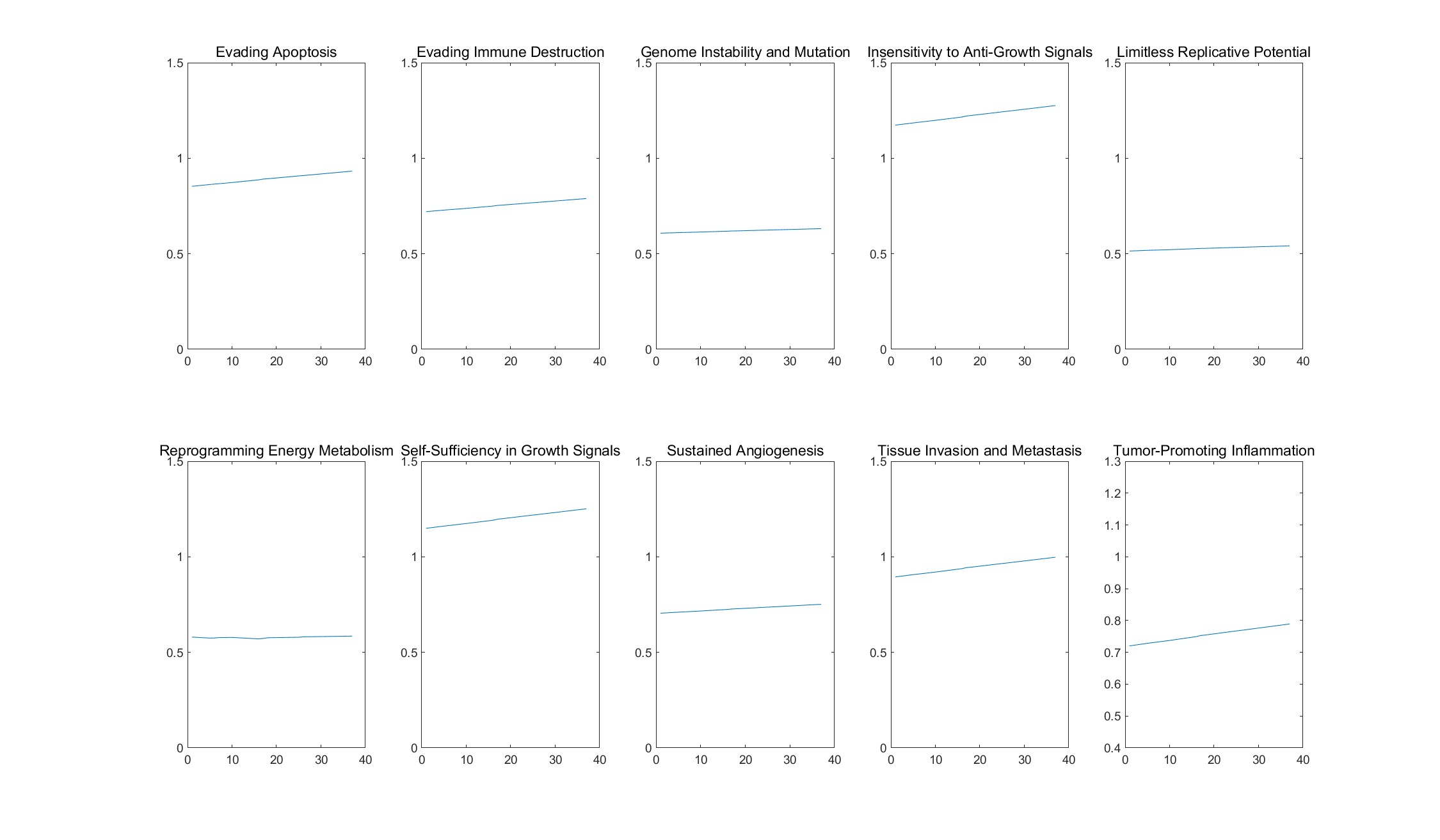}}
    \caption{
      \textbf{(a)} Three-dimensional landscape of cancer evolution.
      \textbf{(b)} Two-dimensional landscape depicting the primary transition path from the normal to the cancerous state.
      \textbf{(c)} Changes in the gene expression levels of the ten hallmark nodes along equidistant steps on the primary transition path.
    }
    \label{fig:potential energy}
\end{figure}

Furthermore, the evolution from the normal to the cancerous state is depicted via landscape and flux theory \cite{Li2014,Lv2024}. In the two-dimensional landscape (Figure \ref{fig:potential energy}), two stable attractors corresponding to normal and cancerous states are identified (Figure \ref{fig:potential energy}a). By tracing equidistant points along the transition path (Figure \ref{fig:potential energy}b), changes in the expression levels of the ten hallmarks can be observed (Figure \ref{fig:potential energy}c). Consistent with previous observations, the most pronounced alterations are observed in \textit{Tissue Invasion and Metastasis}, \textit{Evading Apoptosis}, and \textit{Self-Sufficiency in Growth Signals}. Overall, the pan-cancer analysis demonstrates that network-level reconfiguration precedes and likely drives the subsequent quantitative changes in hallmark expression.

\section{Discussion}
Cancer evolution is a multiscale process encompassing both molecular alterations and emergent system-level properties. A data-driven macroscopic stochastic network evolution framework has been developed to elucidate cancer progression at the systems level. The most important finding is that the reorganization of the inter-hallmark regulatory network is observed to precede the quantitative increases in individual hallmark expression levels. In other words, changes in the network structure are detected before the hallmark nodes exhibit overt abnormal expression, suggesting that the alteration of regulatory connectivity is an early indicator of the malignant transition.

In this work, the pan-cancer analysis conducted across $15$ cancer types from the GRAND database reveals that among the ten hallmarks, \textit{Tissue Invasion and Metastasis} consistently shows the largest divergence between normal and cancerous states. In contrast, hallmarks such as \textit{Reprogramming Energy Metabolism} display only subtle differences, likely because the underlying metabolic pathways are also active in certain normal, rapidly proliferating cells. Furthermore, significant alterations in \textit{Evading Apoptosis} and \textit{Self-Sufficiency in Growth Signals} are observed, underscoring their roles in enabling cancer cells to survive adverse conditions and bypass growth control.

A further analysis using landscape and flux theory demonstrates that the evolution from the normal to the cancerous state can be depicted as a transition between two stable attractors. By tracing equidistant points along the primary transition path, gradual changes in the expression profiles of the hallmarks are quantified. These results, together with the observation that dynamic network biomarker score changes earlier than the expression thresholds, indicate that network reconfiguration is a sensitive early-warning signal than the absolute values such as expression levels.

These findings collectively demonstrate that network-level changes precede phenotypic shifts, suggesting a hierarchical control mechanism in tumor evolution. The results emphasize that alterations in the interactions among hallmarks occur at an early stage of tumorigenesis, potentially offering novel opportunities for early diagnosis and intervention. Pan-cancer analysis confirms that hallmarks such as \textit{Tissue Invasion and Metastasis}, \textit{Evading Apoptosis}, and \textit{Self-Sufficiency in Growth Signals} drive malignant progression, the underlying network-level changes may serve as universal precursors to overt phenotypic shifts in diverse cancer types.

\section{Materials and Methods}\label{methods}

\subsection{Data Collection of Hallmarks}
Gene regulation governs tissue functionality and cellular responses to perturbations. These processes are governed by complex networks of transcription factors, miRNAs, and their target genes, whose architectural organization ultimately determines phenotypic outcomes in health and disease.

The GRAND database \cite{Ben22} provides a comprehensive resource containing 12,468 genome-scale gene regulatory networks, spanning 36 human tissues, 28 cancer types, and 1,378 untreated cell lines. In this study, gene expression data from selected cancer types and both gene expression data and regulatory network data from their corresponding normal tissues were utilized to construct the Hallmark gene regulatory network. The selected cancers include gastric adenocarcinoma, renal clear cell carcinoma, lung adenocarcinoma, and 12 other malignancies, as detailed in Supplementary Table S1.

To establish precise hallmark classifications, Gene Ontology (GO) terms associated with cancer hallmarks were curated from existing literature \cite{Plaisier12}. This collection underwent rigorous refinement through removal of obsolete terms and incorporation of recent updates. Final hallmark gene sets were constructed using the AmiGo tool from the Gene Ontology database \cite{Carbon09}, ensuring consistency with current biological knowledge, as detailed in the Supplementary Materials.

\subsection{Dynamical Equations} 
A network model was constructed where each node represents a cancer hallmark. For each hallmark node, corresponding gene sets were integrated with GRAND database entries, retaining only genes present in both normal and cancer-specific datasets. The expression level of each node was computed as the summation of its constituent genes' expression values, while edge weights between nodes were determined by cumulative regulatory weights of non-overlapping genes (see Supplementary Materials for detailed methodology).

The model incorporates $10$ hallmarks ($M=10$), and $x_i$ ($i=1,2,\cdots,M$) denotes the expression level of hallmark $H_i$, with the system state represented by $x=(x_1,x_2,\cdots,x_M)$. Each hallmark's expression is regulated through bidirectional positive interactions, as captured by the time-varying matrix $V^t$ containing regulatory strengths $v_{ij}(t)$ between hallmark pairs.

The net regulatory effect on hallmark $h_i$ is calculated as:
\begin{equation}
w_i = \sum_{j=1}^{M} \alpha_{ij} V^t_{ij} x_j,
\end{equation}
with uniform weighting coefficients $\alpha_{ij}=1$. The regulatory activation function is defined as:
\begin{equation}
F(w_i) = \rho + (1-\rho) \frac{\sqrt{w_i}}{\theta},
\end{equation}
where $\theta = \mathop{\mathrm{avg}}\limits_{i}\Bigl(\sum\limits_j V^t_{ij}\Bigr)$ and $\rho=0.1$ represents baseline regulatory capacity.

Node dynamics are governed by:
\begin{equation}
\frac{\mathrm{d} x_i}{\mathrm{d}t} = \lambda_i F(w_i) - x_i,\quad (i=1,2,\cdots,M),
\end{equation}
with maximum expression rate $\lambda_i=3.8$. The complete deterministic system is therefore:
\begin{equation}
\begin{array}{rcl}
    \frac{\mathrm{d} x_i}{\mathrm{d} t} &=& \lambda_i F(w_i) - x_i,\quad (i=1,2,\cdots,M), \\
    w_i &=& \sum_{j=1}^{M} \alpha_{ij} V^t_{ij} x_j, \\
    F(w_i) &=& \rho + (1-\rho) \frac{\sqrt{w_i}}{\theta}.
\end{array}
\end{equation}
Key model parameters are summarized in Table~\ref{Table1}.

\begin{table}[ht]
\centering
\caption{Summary of model components and parameters}
\begin{tabular}{|c|p{6.5cm}|}
\hline
Symbol & Explanation \\ \hline
$x_i$ & Expression level of node $H_i$ \\ \hline
$w_i$ & Effective regulatory level of node $H_i$ \\ \hline
$\alpha_{ij}$ & Weighting coefficient \\ \hline
$F(w_i)$ & Regulatory activation function \\ \hline
$v_{ij}$ & Regulatory strength from node $H_i$ to $H_j$ \\ \hline
$\lambda_i$ & Maximum expression rate of node $H_i$ \\ \hline
$\theta$ & Normalization parameter \\ \hline
\end{tabular}
\label{Table1}
\end{table}

To incorporate biological variability, stochastic perturbations were introduced through modification of the degradation rate:
\begin{equation}
\begin{aligned}
\frac{d x_i}{dt} &= \lambda_i F(w_i) - e^{\eta_i - \frac{\sigma^2}{2}} x_i,\quad (i=1,2,\cdots,M), \\
d \eta_i &= -\frac{1}{\tau}\,\eta_i\,\mathrm{d} t + \sqrt{\frac{2}{\tau}}\,\sigma\,\mathrm{d} W_i(t),
\end{aligned}
\end{equation}
where $W_i(t)$ denotes a Wiener process with correlation time $\tau=1$ and noise intensity $\sigma=0.1$. The Ornstein-Uhlenbeck process $\eta_i$ satisfies:
\begin{equation}
E[\eta_i(t)] = 0,\quad E[\eta_i(t_1)\eta_i(t_2)] = \sigma^2 e^{-|t_1-t_2|/\tau}.
\end{equation}

Numerical solutions were obtained using an Euler-Maruyama scheme:
\begin{equation}
\begin{aligned}
x_i^{t+1} &= x_i^t + \Delta t \Bigl(\lambda_i F(w_i^t) - e^{\eta_i^t - \frac{\sigma^2}{2}} x_i^t\Bigr), \\
\eta_i^{t+1} &= \eta_i^t - \frac{1}{\tau}\,\eta_i^t\,\Delta t + \sqrt{\frac{2}{\tau}}\,\sigma\,\sqrt{\Delta t}\,Z_i^{t+1},
\end{aligned}
\end{equation}
where $Z_i^{t+1} \sim \mathcal{N}(0,1)$. For enhanced stability, an implicit difference method was implemented:
\begin{equation}
x_i^{n+1} = \frac{\Delta t\,\lambda_i F(x_i^n) + x_i^n}{1 + \Delta t\,e^{\eta_i^n - \frac{\sigma^2}{2}}}.
\end{equation}

\backmatter

\bmhead{Data Availability} 
The datasets used in the present study are all publicly available. 
The primary data used in this study are available in the GRAND database (\url{https://grand.networkmedicine.org}). 
All the genes in the hallmark-of-cancer related GO terms were downloaded from Gene Ontology (\url{https://geneontology.org/}). 
The list of cancer names selected for this study in the GRAND database is in Supplementary Table S1. 
Detailed information on the GO term names corresponding to Hallmarks and their GO term IDs is provided in Supplementary Table S2. 
Final hallmark gene sets are provided in Supplementary Table S3. 

\bmhead{Code Availability} 
Analysis pipelines and simulation codes are maintained at \url{https://github.com/zhuge-c/Hallmark_dynamics}

\bmhead{Supplementary information}
See supplementary material for details

\bmhead{Acknowledgements}
This work was supported by the National Natural Science Foundation of China (11801020).

\bmhead{Author contributions}
Conceptualization: C. Zhuge, Y. Han, D. Xu; Data curation: Y. Wu, Y. Hou, J. Wang; Computational resources: D. Xu, Y. Li, C. Zhuge; Investigation: J. Wang, Y. Wu, C. Zhuge; Project administration: C. Zhuge; Supervision: D. Xu, Y. Han, C. Zhuge; Writing – original draft: J. Wang, Y. Wu, C. Zhuge; Writing – review and editing: D. Xu, C. Zhuge, Y. Li, Y. Han, J. Wang, Y. Wang, Y. Hou

\section*{Declarations of competing interests}
The authors declare no conflicts of interest.

\section*{Ethical statement}
Not applicable.


\bibliography{ref}

\appendix
\section{Supplemental Material}
\subsection{Data collection}\label{sec:Data collection}
The \textbf{GRAND database} \cite{Ben22} contains data of cancer and tissue datasets, from which 15 cancer types were included in this study. The selected cancers comprise \textbf{gastric adenocarcinoma, renal clear cell carcinoma, lung adenocarcinoma}, and 12 other malignancies as detailed in \textbf{Supplementary Table S1}.
The \textbf{Hallmark gene regulatory network} in this study was constructed using \textbf{gene expression data} from selected cancer types in the \textbf{GRAND database}, along with \textbf{regulatory network data} and \textbf{gene expression data} from their corresponding normal tissues. The datasets used in this study can be accessed based on the cancer and tissue names listed in \textbf{Supplementary Table S1}.

\textbf{Gene Ontology (GO) terms} related to cancer hallmarks were collected from existing literature \cite{Plaisier12}. The set of terms was refined by removing outdated terms and incorporating the most recent updates. Detailed information on the GO terms is provided in \textbf{Supplementary Table S2}.

\subsection{Data processing}\label{sec:Data processing}
Gene Ontology (GO) terms corresponding to each Hallmark were identified through a comprehensive literature review and relevant medical knowledge. Using the \textbf{AmiGO tool} \cite{Carbon09}, gene sets associated with these terms were retrieved from the \textbf{GO Ontology database}. To construct the gene set for each Hallmark, the retrieved gene sets were merged and deduplicated.

Since gene sets from the \textbf{GO Ontology database} are provided as gene symbols, while regulatory network data from the \textbf{GRAND database} use gene IDs as column names and gene symbols as row names, and gene expression data use gene IDs as row names, a standardization step is necessary. To ensure consistency in data integration and facilitate the construction of \textbf{Hallmark-specific gene sets}, it is necessary to convert \textbf{Ensembl gene IDs} into \textbf{HGNC gene symbols}.
This conversion was conducted using the \textbf{``ensembl\_id\_convert.R"} script, which leverages the \textbf{biomaRt} package to retrieve gene annotation data from the \textbf{Ensembl database}. Specifically, the script establishes a connection to the Ensembl database and queries the \textbf{human gene dataset (hsapiens\_gene\_ensembl)}. The corresponding \textbf{HGNC gene symbols} for each Ensembl ID are then extracted using the \textbf{getBM()} function. The final mapped results are systematically stored for subsequent analyses.
For genes that lacked an associated gene symbol and were only available as gene IDs, their gene IDs were assigned as substitutes to maintain dataset consistency. All gene IDs in the \textbf{GRAND database} were successfully mapped to gene symbols for subsequent analysis. Detailed results can be found in \textbf{Supplementary Table S3}.

Discrepancies between the gene sets derived from the \textbf{GO Term database} and those obtained from the \textbf{GRAND database} necessitated further standardization. To ensure consistency and accuracy, the final \textbf{Hallmark gene set} was obtained by computing the intersection of these two gene sets. The detailed composition of these gene sets is provided in \textbf{Supplementary Table S4}.
Using data from the \textbf{GRAND database}, the final \textbf{Hallmark-associated gene set} was constructed, comprising both \textbf{gene expression data} and \textbf{gene regulatory data}.

The expression level of each \textbf{Hallmark node} was computed as the aggregated expression of its constituent genes. The edge weight between two nodes was assigned based on the cumulative regulatory influence of their \textbf{non-overlapping genes}. 
Using these processed data, the \textbf{Hallmark gene regulatory network} was constructed via the \textbf{NetworkConstruct.py} script, serving as the basis for subsequent analyses.
\subsection{Direct interaction network-based divergence}\label{sec:DataDIND}
The early warning method used in this study is based on the \textbf{dynamic network biomarker (DNB)} theory \cite{Chen12}, specifically the computational approach known as \textbf{direct interaction network-based divergence (DIND)} \cite{Peng22}. The detailed methodology can be found in the original references.

Let \( M = 10 \) denote the number of hallmark nodes, and \( x_i \, (i = 1, 2, \dots, M) \) represents the expression level of hallmark \( h_i \). The state of the system is represented as \( x = (x_1, x_2, \dots, x_M) \), and the interaction between nodes is modeled through the time-varying regulatory matrix \( V^t = [v_{ij}^t] \), where \( v_{ij}^t \) denotes the regulatory strength from hallmark \( h_j \) to hallmark \( h_i \). The net regulatory effect on \( h_i \) is calculated as:

\[
w_i = \sum_{j=1}^{M} \alpha_{ij} v_{ij}^t x_j,
\]

where \( \alpha_{ij} \) represents the contribution weight of the interaction between \( h_i \) and \( h_j \), and \( x_j \) is the expression level of hallmark \( h_j \).

To detect critical transitions in the hallmark network, we utilize DIND to quantify the divergence of network states across time. The local divergence between two states is calculated using the Kullback-Leibler (KL) divergence. For two multivariate normal distributions \( N_1 \) and \( N_2 \) with means \( \mu_1, \mu_2 \) and covariance matrices \( \Sigma_1, \Sigma_2 \), the symmetric KL divergence is defined as:

\[
D_{\text{DIND}}(N_1, N_2) = \frac{1}{2} \Big( D(N_1 || N_2) + D(N_2 || N_1) \Big),
\]

where:

\[
D(N_1 || N_2) = \frac{1}{2} \left[ \text{tr}(\Sigma_2^{-1} \Sigma_1) + (\mu_2 - \mu_1)^T \Sigma_2^{-1} (\mu_2 - \mu_1) - d + \ln \frac{\det \Sigma_2}{\det \Sigma_1} \right],
\]

and \( d \) is the dimension of the distributions. The global DIND score for the network at time \( t \) is then computed as:

\[
D_t = \frac{1}{M} \sum_{i=1}^{M} D_t^i,
\]

where \( D_t^i \) is the local divergence score for the \( i \)-th node. A sharp increase in \( D_t \) signals a critical transition in the network, reflecting significant changes in hallmark interactions.

\subsection{Potential}\label{sec:Potential}
Given a matrix $X \in \mathbb{R}^{n \times m}$, where $n$ is the number of samples and $m$ is the number of features, the goal of PCA is to perform an orthogonal transformation of the data, finding a new coordinate system where the variance of the data is maximized. Specifically, PCA is carried out through the following steps:

1. Centering the Data

First, we center the data matrix $X$ by subtracting the mean of each feature (column), so that each feature has a mean of zero.

\[
X_{\text{centered}} = X - \text{mean}(X)
\]

2. Covariance Matrix Calculation

Next, compute the covariance matrix $\Sigma$ of the centered data, which describes the linear relationships between different features.

\[
\Sigma = \frac{1}{n-1} X_{\text{centered}}^T X_{\text{centered}}
\]

3. Eigenvalue and Eigenvector Calculation

Perform an eigenvalue decomposition of the covariance matrix $\Sigma$ to obtain the eigenvalues and eigenvectors. The eigenvectors correspond to the directions of the principal components, and the eigenvalues represent the variance captured by each principal component.

\[
\Sigma v = \lambda v
\]

where $v$ is an eigenvector and $\lambda$ is the corresponding eigenvalue.

4. Selecting Principal Components

Select the top $k$ eigenvectors (principal components) based on the size of the eigenvalues. These correspond to the directions of maximum variance in the data.

5. Projecting the Data

Project the original data onto the selected principal components to obtain a new representation of the data, $Z$, where each column represents the coordinates of the data in the principal component space.

\[
Z = X_{\text{centered}} V_k
\]

where $V_k$ is the matrix of the top $k$ eigenvectors.

In the study, we select 1000 samples from normal and cancer respectively, and the 10 hallmarks as features to performs the PCA process. The function will output the following:
  \begin{itemize}
      \item \textbf{Principal Component Matrix}: A $10 \times 10$ matrix where each column is an eigenvector representing the new principal component direction.
      \item \textbf{Variance (Eigenvalues)}: A vector of size 10 containing the variance (eigenvalues) associated with each principal component, representing their importance.
      \item \textbf{Projected Data}: A $2000 \times 10$ matrix representing the data projected into the principal component space.
  \end{itemize}

In order to depict the change of Hallmarks from normal to cancer, Kernel Density Estimation (KDE) is used to calculate the changes in the potential functions of the first two principal components.

\begin{equation}
\operatorname{potential}\left(x_j, y_j\right)=\frac{1}{n \cdot h \sqrt{2 \pi}} \sum_{i=1}^n \exp \left(-\frac{\left(x_i-x_j\right)^2+\left(y_i-y_j\right)^2}{2 h^2}\right),\notag
\end{equation}
where $n$ is the number of data points, $h$ is the bandwidth argument that controls the width of the kernel function (in the study, the bandwidth is 0.5).

\end{document}